\documentclass[sort&compress,english,preview,5p]{elsarticle}
\journal{Ultramicroscopy}

\usepackage{graphicx}
\usepackage{xcolor}
\usepackage{tabularx}
\usepackage{amsmath}
\usepackage{amssymb}
\usepackage{physics}
\usepackage{appendix}
\usepackage{comment}

\usepackage{natbib}

\begin{document}
\title{Automated detection and mapping of crystal tilt using thermal diffuse scattering in transmission electron microscopy}

\author[fzj]{Mauricio Cattaneo}
\author[lmu,fzj]{Knut Müller-Caspary\corref{cor1}}
\author[fzj]{Juri Barthel\corref{cor1}}
\author[fzj]{Katherine~E. MacArthur} 
\author[emat]{Nicolas Gauquelin} 
\author[fzj,gfe]{Marta Lipinska-Chwalek}
\author[emat]{Johan Verbeeck} 
\author[uom]{Leslie~J. Allen} 
\author[fzj]{Rafal~E. Dunin-Borkowski}

\cortext[cor1]{To whom correspondence should be addressed by E-mail to:\\
k.mueller-caspary@cup.lmu.de (K.\,M.-C.)\\
ju.barthel@fz-juelich.de (J.\,B.)}

\address[fzj]{Ernst Ruska Centre (ER-C), Forschungszentrum J{\"u}lich GmbH, 52425 J{\"u}lich, Germany}
\address[lmu]{Department of Chemistry and Centre for NanoScience, Ludwig-Maximilians-University Munich, Butenandtstr. 11, 81377 Munich, Germany}
\address[emat]{Electron Microscopy for Materials Science (EMAT), University of Antwerp, 2020 Antwerp, Belgium}
\address[uom]{School of Physics, University of Melbourne, Parkville, Victoria 3010, Australia}
\address[gfe]{Central Facility for Electron Microscopy, RWTH Aachen University, 52074 Aachen, Germany}

\begin{abstract}
Quantitative interpretation of transmission electron microscopy (TEM) data of crystalline specimens often requires the accurate knowledge of the local crystal orientation. A method is presented which exploits momentum-resolved scanning TEM (STEM) data to determine the local mistilt from a major zone axis. It is based on a geometric analysis of Kikuchi bands  within a single diffraction pattern, yielding the centre of the Laue circle. Whereas the approach is not limited to convergent illumination, it is here developed using unit-cell averaged diffraction patterns corresponding to high-resolution STEM settings. In simulation studies, an accuracy of approximately 0.1~mrad is found. The method is implemented in automated software and applied to crystallographic tilt and in-plane rotation mapping in two experimental cases. In particular, orientation maps of high-Mn steel and an epitaxially grown La$_{\text{0.7}}$Sr$_{\text{0.3}}$MnO$_{\text{3}}$-SrTiO$_{\text{3}}$ interface are presented. The results confirm the estimates of the simulation study and indicate that tilt mapping can be performed consistently over a wide field of view with diameters well above 100\,nm at unit cell real space sampling.
\end{abstract}
\maketitle
\section{Introduction}
Scanning transmission electron microscopy (STEM) has established itself as a method for structural characterization as well as chemical composition mapping with atomic-scale resolution. A diversity of imaging techniques has been developed, such as annular dark field (ADF) imaging providing $Z$-contrast \cite{pennycook1988zcontrast}, bright field (BF) and annular bright field (ABF) imaging of light elements \cite{findlay2010dynamics}, elemental mapping using electron energy-loss spec\-tros\-co\-py (EELS) \cite{bosman2007atomiceels} and energy-dispersive x-ray spec\-tros\-co\-py (EDX) \cite{dalfonso2010atomicedx}.

Common to these methods is a signal dependence on the crystallographic orientation of the specimen with respect to the incident electron beam. Even relatively small mistilts away from a crystal zone axis can affect the dynamical scattering of the electron wave significantly \cite{maccagnano2008effects, Zhou2016, macarthur2021optimizing}. Quantitative conclusions can often be drawn only via the comparison of experimental data with accompanying simulations, which require  accurate tilt information as input to model scattering dynamics realistically. In the case of EDX spectroscopy, for example, a deliberate sample tilt is applied in order to suppress electron channelling~\cite{MacArthur2017, Lugg2015}. Furthermore, structure retrieval by, e.g., inverse multislice~\cite{Chen2021,Sha2023,Diederichs2024,Broek2012} can benefit either from the knowledge of local tilt as initial input, or as a sanity check of the reconstruction result.

Furthermore, research partly focuses on the local tilt distribution itself. Examples include materials where local variations of the orientation due to, e.g., stacking faults, interfaces, dislocations and grain boundaries are related to materials properties such as elasticity or ferroelectricity~\cite{liao2018,calcagnotto2010,Strauch2023}. A technique to measure and map the local sample tilt would thus be beneficial for such applications, in particular if the  data providing tilt information can be acquired simultaneously with, e.g., spectroscopic or Z-contrast data. In this regard, the advent of momentum resolution in STEM using ultra-fast cameras \cite{Muller2012a,Plackett2013,Muller-Caspary2015a,Ryll2016,Tate2016,Jannis2022} has dramatically expanded the versatility and flexibility of STEM, as complete diffraction patterns can be acquired at each scan position up to a certain spatial frequency. In that respect,  four-dimensional STEM (4D-STEM) data sets have already been used to determine specimen tilt and thickness in an automated way by means of neural networks analyzing the elastic scattering signal \cite{xu_lebeau_cnn}. Moreover, a correlative approach comparing mainly the elastic scattering signal in experimental and simulated position-averaged convergent beam electron diffraction (PACBED) patterns has also shown a high sensitivity to sample mistilt \cite{lebeau_pacbed_2010}. Because these approaches are crystal and zone axis specific, they require prior knowledge. In the method presented here, the amount of needed prior information is significantly reduced while maintaining a high evaluation speed and accuracy.

At least for specimens investigated at sufficiently large thicknesses and temperatures, typically above 10\,nm at 300\,K, thermal diffuse scattering (TDS) constitutes a considerable fraction of the dark-field intensity. Besides the smooth diffuse background, Kikuchi bands can be observed in diffraction patterns of crystals at high scattering angles~\cite{nishikawa1928diffraction}. Since Kikuchi bands are aligned with reciprocal lattice planes, their exploitation to measure crystal tilt is widespread in electron diffraction. For example, crystallographic orientation mapping via Kikuchi bands has a long tradition in scanning electron microscopy (SEM) using electron back-scatter diffraction (EBSD)~\cite{Venables1973,trimby_kik_sem,burton_Kik_EDAX}. The kernel of the existing Kikuchi band detection schemes in EBSD usually involves a Hough transform to locate the Kikuchi band crossings. However, Hough-based methods can become imprecise in cases where rectangular-shaped Kikuchi bands do not constitute the dominant features in diffraction patterns. A strong zero-order Laue zone in diffraction patterns recorded in transmission significantly complicates the application of Hough transforms, especially at high beam energies in the (S)TEM. 

In this work, we present an alternative approach based on Kikuchi band recognition from azimuthal intensity profiles in an annular dark-field region of a diffraction pattern. While requiring only a minimum of input, such as the expected number of strong Kikuchi bands, the method performs very robustly in typical high-resolution STEM scenarios in, or close to, zone-axis orientation and allows us to quantify mistilt with well below 1~mrad accuracy and precision.  Whereas, methodologically, each diffraction pattern is treated independently, the application of this scheme also provides the ability to map local tilts in large 4D STEM data sets. The procedure was implemented in a computer program to run fully automatically and computationally efficiently, such that the detection of the centre of the Laue circle takes on the order of several milliseconds per diffraction pattern on current personal computer hardware.

In the following, the method to measure the tilt between a crystal axis and the incident beam direction based on single diffraction patterns is introduced conceptually. Major steps of the analysis include the enhancement of Kikuchi band contrast in the dark-field region using a renormalization scheme, the identification of Kikuchi band crossings from azimuthal profiles, as well as iteratively eliminating systematic errors arising from the initially unknown centre of the Laue circle. Frozen phonon multislice simulation studies are performed in order to assess accuracy and precision of the method in dependence on electron dose and mistilt. The method is then applied to two experimental cases demonstrating tilt mapping (1)~across a large field of view spanning more than 100\,nm in a bent cross-section containing an La$_{\text{0.7}}$Sr$_{\text{0.3}}$MnO$_{\text{3}}$/SrTiO$_3$ interface, and (2)~to a smaller scale of approximately 10\,nm across a twin boundary in high-Mn steel. For both examples, 4D-STEM data sets were recorded with atomic column resolution. Before the tilt analysis, the diffraction patterns were averaged over scan areas corresponding to the size of one projected unit cell. Mistilt maps are thus generated from PACBED patterns with unit cell sampling. Due to the robustness and accuracy of the method, small changes of sample tilt are resolved over a large field of view as well as with high sampling close to interfaces and grain boundaries.

\section{Methods}

\subsection{Conceptual outline and problem statement}
The measurement of sample mistilt from a zone axis orientation based on Kikuchi bands, as proposed here, consists of locating the crossing point of the bands by geometrical construction and determining its distance from the center of the bright-field disc. In principle, this analysis could be performed exploiting the full information on tilt present in a two-dimensional diffraction pattern, e.g., by cross-correlation against an expected band pattern. However, the intensity distribution and details like the shape and width of Kikuchi bands depend on the crystal structure, sample thickness, zone-axis orientation, and further experimental parameters. This means a full two-dimensional analysis of the Kikuchi-band pattern obtained in transmission electron diffraction which would require extensive information as input to an automated routine. Since we are aiming for a fast method which requires only a minimum of prior knowledge, the evaluation approach is simplified by determining points in the diffraction plane through which Kikuchi bands pass.


\begin{figure}
\centering
\includegraphics[width=\linewidth]{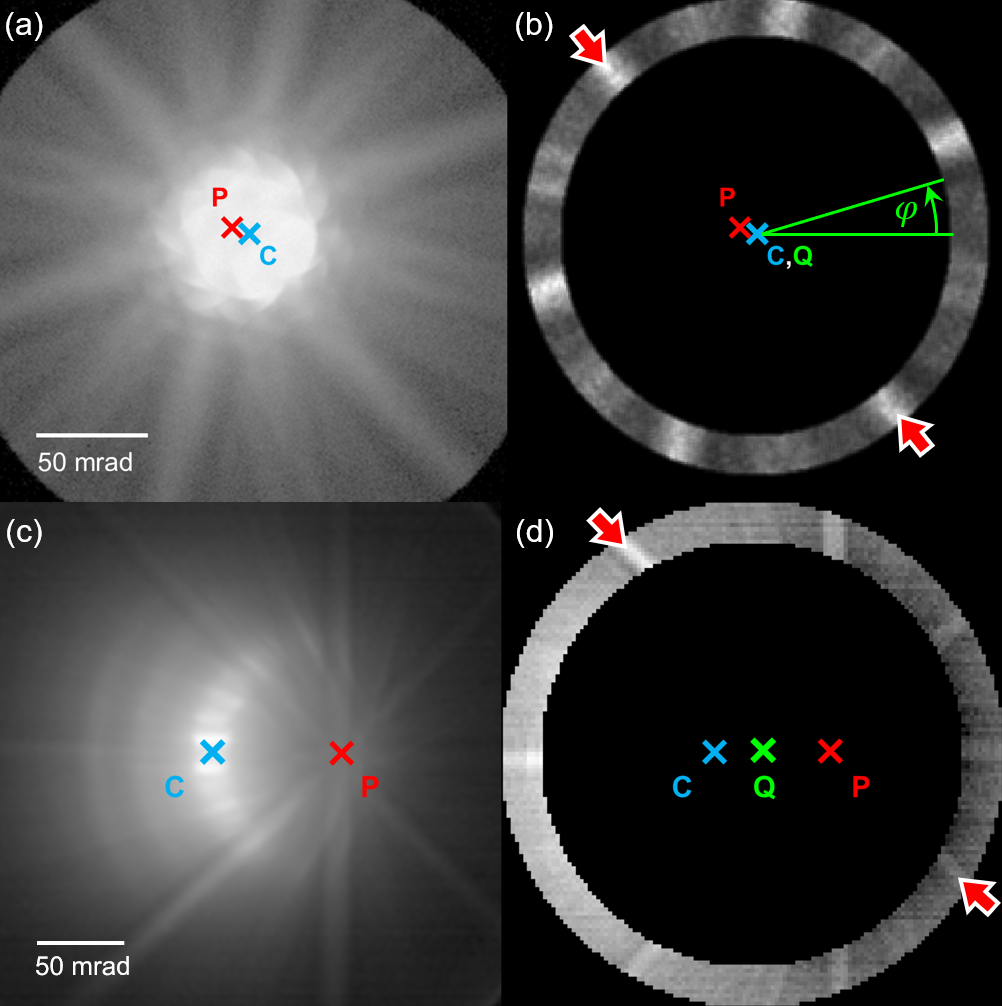}
\caption{(a) Diffraction pattern of a steel sample recorded with 300\,keV electrons and at 7.2\,mrad mistilt from the [110] zone axis. Kikuchi bands intersect in point $\bf{P}$ whereas the incident beam direction corresponds to point $\bf{C}$. The logarithm of intensity is plotted to enhance the visibility of Kikuchi bands. (b) Re-normalized signal of the pattern in (a) within an annular region that is centered around the point $\bf{Q} = \bf{C}$. (c) Diffraction pattern of a SrTiO$_3$ sample recorded with 300-keV electrons at a large mistilt of about 75 mrad from the [100] zone axis. (d) Signal of the pattern in (c) within an annular region around $\bf{Q}$ in the detector center. Arrows indicate two traces of one Kikuchi band in the annular mask.}
\label{renorm_scheme_largetilt}
\end{figure}

In a preliminary step, an annular mask is applied to the diffraction pattern so as to separate the data including thermal diffuse scattering (TDS) and thus Kikuchi bands at medium to high angles from the bright field and low-angle diffraction. An example with a relatively small mistilt of 7.2\,mrad is shown in Fig.~\ref{renorm_scheme_largetilt}\,(a,b) for the full and masked experimental diffraction pattern, respectively. In the following, we denote the locations of the undiffracted beam by $\mathbf{C}$, the centre of the Laue circle by $\mathbf{P}$ and the centre of the annular mask by $\mathbf{Q}$. As indicated by the arrows in Fig.~\ref{renorm_scheme_largetilt}\,(b), Kikuchi bands passing through the mask leave two traces at approximately opposite sides of the annulus with azimuths differing by approximately $\pi$. Figure~\ref{renorm_scheme_largetilt}\,(c,d) shows an example with a comparably large mistilt of 75\,mrad, and a ring mask that is neither centered around $\mathbf{C}$ nor $\mathbf{P}$. In such cases, the azimuth difference corresponding to a single Kikuchi band can deviate significantly from $\pi$. Moreover, the centering of the mask away from $\mathbf{C}$ leads to a strongly varying mean scattered intensity in dependence of the azimuth~$\varphi$, and the traces of two bands can become very close and might even overlap in some cases. In addition, there can be a much higher intensity close to the inner rim of the mask. 

These circumstances complicate an automatic and robust detection of Kikuchi bands by determining the azimuthal positions of their traces in the annular mask. Furthermore, the signal itself decreases along the radial direction, i.e. towards larger scattering angles where it approaches a Rutherford type of decay of the scattering cross-section \cite{reimer_tem}. This would inherently increase the weight of scattering close to the inner rim of the annulus if not corrected for during the evaluation. In that respect, it is noteworthy that the azimuthal resolution increases with the distance from the mask center. In the following, solutions to  these problems are described which provide an enhancement of the Kikuchi-band signal, facilitate locating band positions within the annular region, and finally enable the determination of the position of the common band crossing, that is, the centre of the Laue circle.

\subsection{Enhancing Kikuchi band contrast}
Since Kikuchi bands are not a dominant feature in the bright field and also in the low-angle dark-field region, a short camera length is beneficial for accessing large scattering angles such that the Bragg reflections within the zero-order Laue zone in the center cover less than a quarter of the recorded diffraction pattern, as shown by the examples in Fig.~\ref{renorm_scheme_largetilt}. For a given camera length and diffraction projection on the pixelated detector, we assume that both Kikuchi-band crossing $\mathbf{P}$ and bright field $\mathbf{C}$ are inside the inner radius of the annular mask. Consequently, the geometrical approach developed in the following is restricted to tilt variations limited by the inner radius of the annular mask at the present stage. However, since this already covers several tens of milliradians tilt variation, it is applicable to the majority of practically relevant cases. The outer radius of the mask is made as large as possible reaching out to the edge of the detector area.


Before an azimuthal scan is extracted from the area selected by the annular mask, the mask is partitioned in concentric rings of widths between one and two detector pixels. For each thin annular partition, the mean value of intensity is subtracted and the remaining signal variation is normalized with respect to the standard deviation. Each partition is then of equal strength in signal variation. For the subsequent integration of the signal along the radial direction, this renormalization avoids stronger weight of inner parts of the annular mask compared to outer parts due to the decay of the electron scattering cross-section towards larger angles. The result of renormalization within the annular mask is shown in Figs.~\ref{renorm_scheme_largetilt}\,(b) and (d), where the Kikuchi bands show only weak intensity variation along the radial direction, but a strong variation occurs azimuthally.

\subsection{Azimuthal scan and Kikuchi band detection}
A one-dimensional azimuthal intensity profile depending on $\varphi$ is extracted from the annular mask by integrating the renormalized data along the radial direction. Sufficient azimuthal sampling is given, for example, by the number of pixels crossed by the inner circle limiting the annular mask. For each azimuthal sample the intensity is effectively integrated over a larger azimuthal range (1)~in order to suppress the fine structure of Kikuchi bands which could possibly impede subsequent peak finding and (2)~to further improve signal-to-noise ratio. The size of these azimuthal sectors should be close to the size of the Kikuchi bands within the annular mask. In that respect, sectors spanning five to ten degrees were found to be good choices in all cases analyzed so far.

\begin{figure}
\centering
\includegraphics[width=0.99\linewidth]{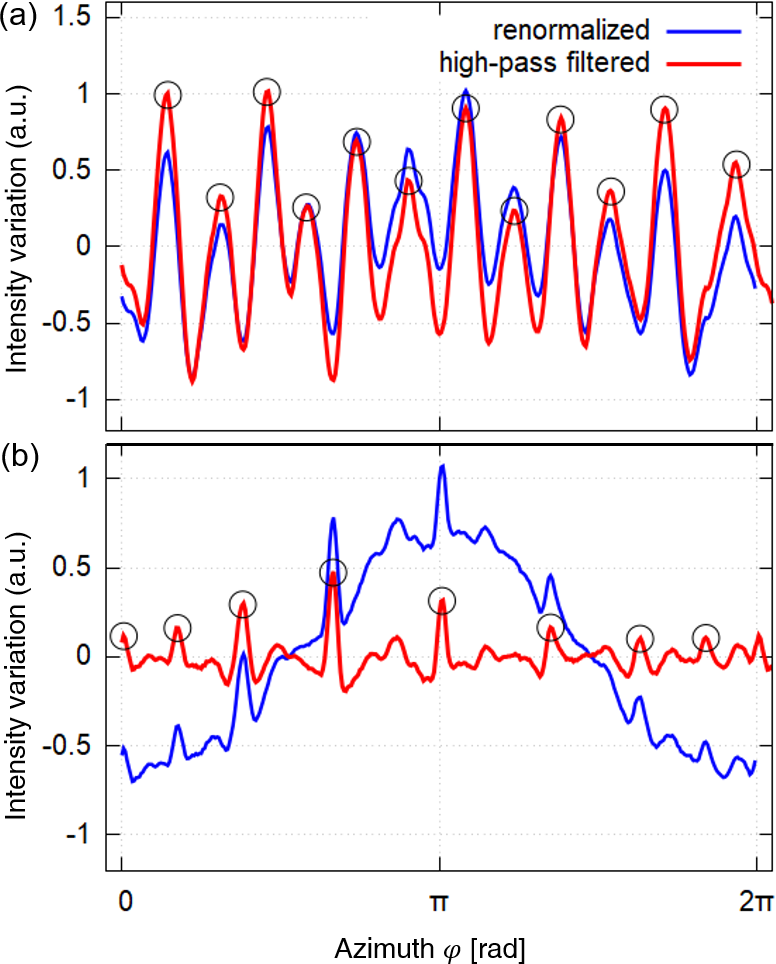}
\caption{Azimuthal scan (a) as extracted from the [110] steel diffraction pattern in Fig.~\protect\ref{renorm_scheme_largetilt}(b) and (b) as extracted from the [100] SrTiO$_3$ pattern in Fig.~\protect\ref{renorm_scheme_largetilt}(d). The azimuthal sampling was done in steps of one degree and binning into sectors spanning six degrees (blue curves). Kikuchi bands appear as narrow peaks in the profile. A strong background modulation is present because parts of the azimuthal scan are closer to the bright field than others. This background modulation is removed in the red curves by means of a high-pass filter. Black circles indicate the result of local peak detection.}
\label{freq_filter_azimuthal}
\end{figure}

Figure~\ref{freq_filter_azimuthal}\,(a) shows the azimuthal scan that has been directly extracted from the pattern in Fig.~\ref{renorm_scheme_largetilt}\,(b) in blue, using sectors with a width of six degrees. The three major and three minor Kikuchi bands in this pattern add up to twelve peaks in total. The azimuthal positions of the peaks are measured by detecting the local maxima, and the mean radius of the annular mask is used as radial position.

The azimuthal profile in Fig.~\ref{freq_filter_azimuthal}\,(b) has been extracted directly from the example with larger mistilt in Fig.~\ref{renorm_scheme_largetilt}(d). It shows a strong one-fold, low-frequency modulation with an amplitude similar to that of the set of peaks caused by Kikuchi bands which impedes robust automatic peak detection and stems from the large deviation of $\mathbf{Q}$ from $\mathbf{C}$. Such a low-frequency modulation is also present in the blue curve of Fig.~\ref{freq_filter_azimuthal}(a), although with a much lower amplitude. A higher intensity occurs for azimuths where the mask is closer to the bright field and covers lower scattering angles. Besides  difficulties in automated peak detection, this effect causes a significant shift of peaks located at strong slopes of the background. In order to prevent  systematic errors arising from this effect, a high-pass filter is applied to the azimuthal scan which effectively removes the low-frequency background. This high-pass filter sets the lowest-order Fourier coefficients to zero resulting in the red profiles in Fig.~\ref{freq_filter_azimuthal}, which are now dominated by the peaks caused by Kikuchi bands in terms of signal variation.

The black circles in Fig.~\ref{freq_filter_azimuthal}(a) indicate the positions of the twelve strongest local maxima of the red curve expected for six Kikuchi bands in the case of a steel sample investigated close to [110] orientation. Likewise, the eight highest local maxima occurring in the red curve of Fig.~\ref{freq_filter_azimuthal}(b) mark the locations expected from four strong Kikuchi bands for [100] SrTiO$_3$. In the example of Fig.~\ref{freq_filter_azimuthal}\,(a), peaks have comparable heights, stand out significantly from the background and can hence be detected straight forwardly by established methods. However, the situation is different for the high-pass filtered azimuthal profile in Fig.~\ref{freq_filter_azimuthal}\,(b) which shows a large number of local maxima at azimuths not related to the centers of Kikuchi bands. A classification of a local maximum as a Kikuchi band or as a peak caused by noise related to, e.g., low counting statistics, is not obvious even for a human observer at least in some cases. For example, setting a threshold based on peak height, width, area or any other quantity regarded as peak strength, may easily reject actual Kikuchi-bands while accepting noise-related peaks. Therefore, the detection routine is allowed to find more than the expected number of local maxima and registers them in a sorted list by means of a parameter of strength, e.g. by the peak height. This approach allows for the identification of falsely detected peaks in the next step, based on a criterion that leads to the expected geometrical situation of Kikuchi bands, namely that they are crossing in one point of the diffraction pattern. Finally, note the regular spacing of peaks in Fig.~\ref{freq_filter_azimuthal}\,a due to the nearly coinciding centres $\mathbf{P}$ and $\mathbf{Q}$ in Fig.\ref{renorm_scheme_largetilt}\,b, as well as the distorted spacings of maxima on the $\varphi$ axis in Fig.~\ref{freq_filter_azimuthal}\,b where the mask centering deviates significantly from the centre of the Laue circle according to Fig.~\ref{renorm_scheme_largetilt}\,d.

\subsection{Determination of the Kikuchi band intersection}\label{sec_kbintersec}
\label{cluster_section}
\begin{figure}
\centering
\includegraphics[width = 0.95\linewidth]{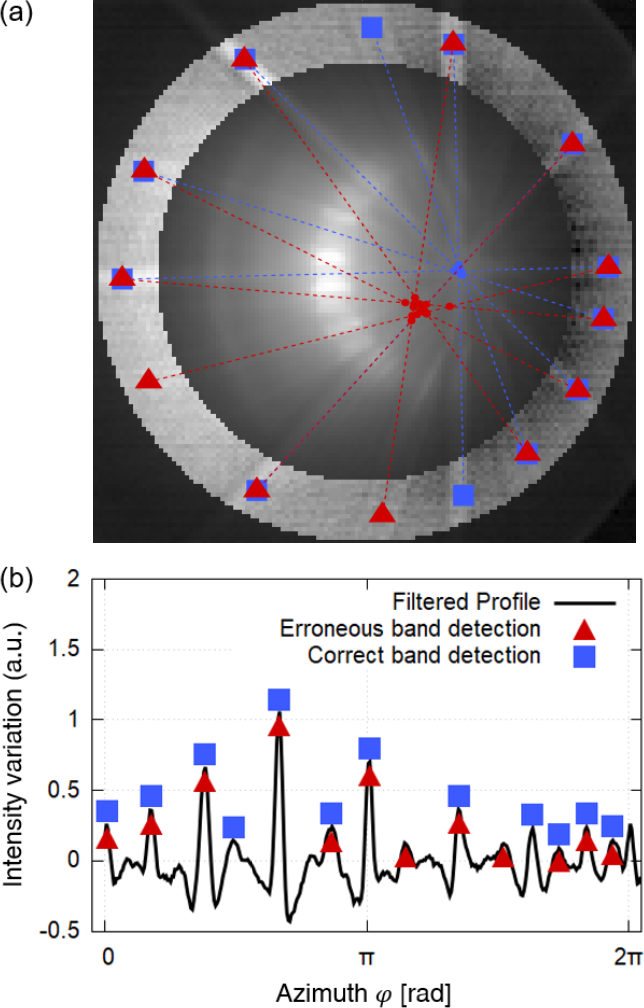} 
\caption{(a) PACBED pattern of [110] fcc steel with larger mistilt and (b) azimuthal scan demonstrating an erroneous (red) and a correct (blue) association of detected peaks to pairs, which are assumed to represent the two traces of Kikuchi bands in the annular mask. The pairs deduced from the list of peaks marked by red triangles lead to the wrong result because the second trace of two Kikuchi bands is not included. The resulting set of crossing points shows a large scatter and is far away from the crossing point of the Kikuchi bands. The list of peaks marked by blue squares leads to a set of crossing points with low scatter at the crossing point of the Kikuchi bands.}
\label{first_step_demo}
\end{figure}

For $n$ Kikuchi bands in a diffraction pattern the annular mask is ideally crossed twice by each band, generating $2n$ peaks in the azimuthal scan. In such cases, a straightforward approach to associate peak pairs to Kikuchi bands, is to connect one peak with the $n^\mathrm{th}$ peak in the azimuthal sequence by a straight line. Thus $n$ lines are constructed which have in total
\begin{equation}
\eta = \frac{n(n-1)}{2}
\label{eq:eta}
\end{equation}
crossings, of which all coincide at or close to the centre of the Laue circle in the ideal situation, as shown by the blue markers in Fig.~\ref{first_step_demo}. The procedure to calculate the crossing point for two pairs of different azimuths is detailed in \ref{AppCalcCrossing}.

This procedure relies entirely on the correct detection of all peaks and, in particular, of the complete set of pairs of peaks related to the Kikuchi bands. The association of peaks to pairs based on the azimuthal sequence fails to produce a set of crossing points forming a compact cloud, if Kikuchi bands are only partially detected. For example, this can happen if only one of the two intersections of a band with the annular mask is found, if two bands overlap and produce only one instead of two peaks, or when a false recognition occurs due to detecting peaks arising from noise. An example for a false relation of points to Kikuchi bands is shown by the red markers in Fig.~\ref{first_step_demo}, where only one of the two intersections of weak bands with the annulus were included. This leads to a false sequential association of pairs and thus to a wide scatter of crossing points quite far away from the common crossing of Kikuchi bands. The residual scatter of the crossing points for a correct selection of peaks is smaller and expected to be in the range of the width of Kikuchi bands. 

In most cases, the set of the $2n$ strongest peaks found in the azimuthal scan provides also the best solution, especially when there are $n$ Kikuchi bands that are clearly visible and sample mistilt is low. In order to stabilize the solution for more challenging scenarios, as the one shown in Fig.~\ref{first_step_demo}, more than $2n$ peaks are taken into consideration. In particular, the list of peaks found in the azimuthal scan is sorted with respect to the peak height, and extended to include also weaker ones which are only potentially related to Kikuchi bands. Several sets of $2n$ peaks are created from this extended list by randomly exchanging mostly weaker peaks. From these sets, the one leading to the lowest scatter of crossing points is chosen as the solution to identify the location of the Kikuchi-band crossing.


\subsection{Iterative error correction}\label{sec_iter}
\begin{figure}
\centering
\includegraphics[width = \linewidth]{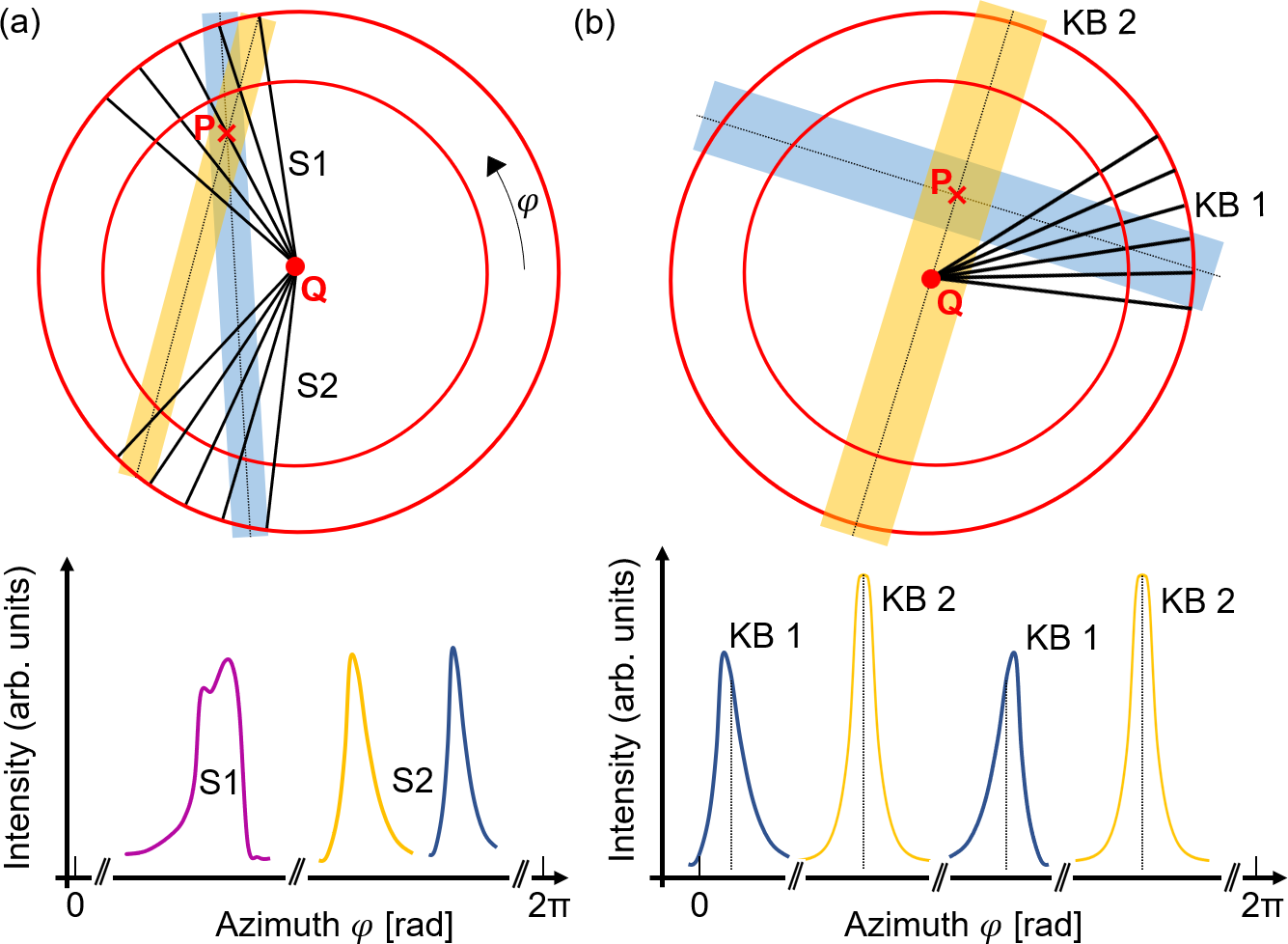} 
\caption{Illustration of systematic errors in Kikuchi band detection due to an off-centered annular mask or significant sample mistilt. (a)~Situation showing two bands crossing under a small angle close to the edge of the annular mask, which makes it difficult to resolve their intersections with the annulus in region ``S1'', whereas two well separated band sections occur in the opposite region ``S2''. (b)~Situation with two perpendicular Kikuchi bands, where the band labelled ``KB 1'' is shifted off the mask center along band ``KB 2''. Symmetric peaks are registered in azimuthal sectors for ``KB 2'', which crosses the annular mask along the radial direction. In contrast, the peaks due to ``KB 1'' are asymmetric, which leads to a systematic shift of the deduced crossing point.}  
\label{large_tilt_distortion}
\end{figure}

On startup of the evaluation procedure, the position $\bf P$ of the centre of the Laue circle in a diffraction pattern is unknown. Especially in the case of large sample mistilt, i.e. for a large distance between points $\bf P$ and $\bf C$ in Fig.~\ref{renorm_scheme_largetilt}, the annular mask applied in the previous steps will usually be centered around a point $\bf Q$ in the diffraction pattern that may be rather remote from the initially unknown point $\bf P$. Consequences are that (1)~Kikuchi bands are more difficult to resolve in parts of the mask close to $\bf P$ and (2)~some bands cause peaks with an asymmetric shape in the azimuthal profile. The first issue can lead to a serious lack of detected pairs, whereas the second one causes a small systematic shift of the result towards the center of the mask. The two problems are illustrated schematically in Fig.~\ref{large_tilt_distortion} with two bands and a significant offset of their crossing from the center position of the annular mask. Errors due to these issues can be quite significant, i.e. larger than the scatter of crossing points found in the previous step.

The two problems described above are minimum if the annular mask is centered at the crossing of Kikuchi bands. Therefore, an iterative scheme is applied in which the annular mask is centered to the result of the previous evaluation and a new evaluation is performed. In most cases only one or two iterations are needed until the correction of the mask position is smaller than the scatter of crossing points.

\subsection{Software implementation}
The evaluation procedure described above has been implemented in a computer program using the \textsc{C++} language. With commercial hardware, each iteration of finding a Kikuchi intersection for a single PACBED takes on the order of ten milliseconds using a single CPU thread. Parallel programming via \textsc{OpenMP} is applied to further speed up the analysis making use of the fact that the evaluation of multiple diffraction patterns can be performed independently. A set of 1000 diffraction patterns is evaluated by four threads in less than 25 seconds.

The program can take one or multiple diffraction patterns as input in form of two-dimensional arrays of intensity values together with parameters describing the size of the array. The inner radius limiting the annular mask and the number of expected Kikuchi bands $n$ are the required input ensuring stability of the procedure. Defining the outer radius of the annular mask is optional but limiting the width of the annulus has shown improved performance in some cases. Output is the position of the Kikuchi band intersection $P_{N}$ in the input array as determined after $N$ iterations and an error estimate for the position corresponding to the root-mean-square deviation of all crossing points of detected Kikuchi bands. The mistilt is then calculated as the distance of the output position $P_{N}$ to the known position $C$ of the incident probe direction considering scaling parameters which describe the diffraction projection in physical units, e.g. in milliradians per pixel.

\section{Simulation studies}\label{sec_simstudy}

In order to assess the accuracy and precision of the mistilt measurement, diffraction patterns were simulated using the multislice method \cite{cowley1957scattering} as implemented in the \textsc{Dr. Probe} software \cite{barthel_drprobe_2018} for a large range of mistilts. Thermal diffuse scattering was simulated according to the quantum excitation of phonons (QEP) model \cite{Forbes2010} assuming an Einstein model of uncorrelated thermal vibration. As a test case, diffraction patterns were calculated for 40~nm thick fcc Fe viewed along the $[110]$ zone axis based on the model of Straumanis \& Kim \cite{straumanis1969lattice}. An aberration-free incident electron probe of 200\,keV kinetic energy and 25\,mrad convergence semi-angle was assumed. PACBED patterns were obtained by summation of patterns resulting from more than 100 probe positions distributed uniformly over the projected unit cell. The diffraction patterns extend out to scattering angles of about 250 mrad due to a fine sampling of the supercell of $5 \times 7$ unit cells with $800 \times 800$ points perpendicular to the incident beam direction. The grid $x$-axis was along the $[\bar{1}10]$ direction and its $y$-axis along $[001]$. This setup also realises a step size of approximately 1\,mrad in the diffraction patterns, which is on the order of magnitude that can be achieved with fast pixelated direct electron detectors and convenient camera lengths in experiments. The potentials were projected along $[110]$ within volume slices that are thin enough to contain just one atomic layer. For each layer 100 positional configurations were calculated by including random displacements with mean squared amplitudes of $\langle u^2 \rangle = 0.007\, \text{\AA}^2$\cite{Peng_DW_factors}. By random selection of positional configurations from the pre-calculated set for each slice during the propagation of the probe function through a thick crystal in the multislice scheme, the scattering for each probe position is essentially calculated with an individual positional configuration. The summation of patterns from many probe positions form a PACBED pattern which thus also includes a summation over positional configurations achieving convergence for the signal related to thermal diffuse scattering. A pattern calculated in this way shows three strong Kikuchi bands along the main reciprocal lattice plains and weaker bands in between.

Test mistilts were distributed uniformly over a range up to 40~mrad tilt magnitude. Patterns obtained directly from the simulations correspond to infinite dose conditions. Scenarios representing a finite dose were simulated by adding noise following Poisson statistics. The total dose assumed in the simulation of one PACBED pattern translates to incident probe currents where we assume here a dwell time of 1 ms per probe position and 100 probe positions contributing to each simulated PACBED. 

\begin{figure}
\includegraphics[width=0.95\linewidth]{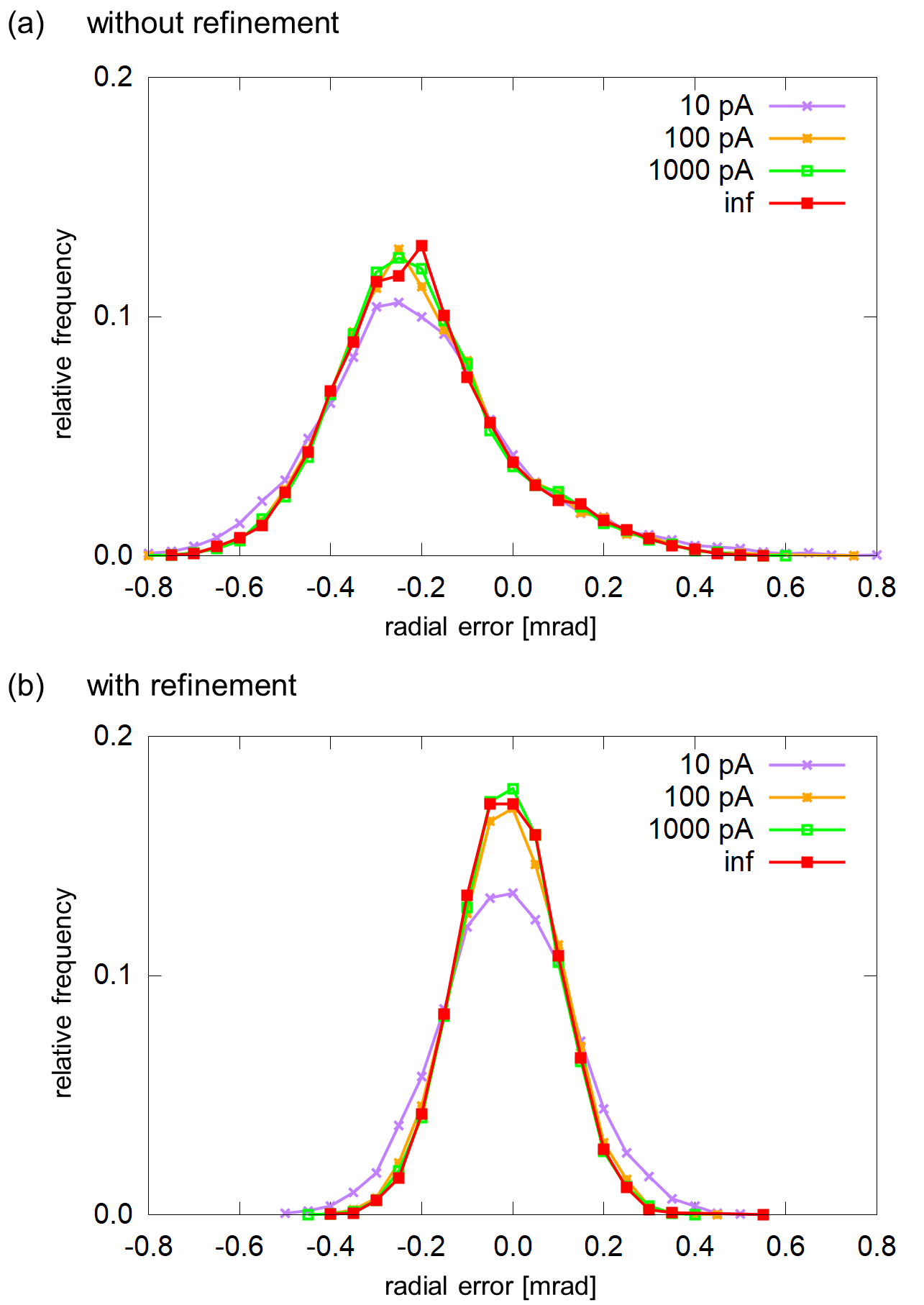}
\caption{Histograms of the overall radial error of mistilt measurements with simulated patterns for different probe currents. (a)~Detection error for a direct evaluation, without iterative refinement. A systematic error of approximately 0.1~mrad is observed. (b)~Improved accuracy after iterative refinement showing no significant systematic error.}
\label{Accuracy_w_wo_it}
\end{figure}

The histograms in Fig.~\ref{Accuracy_w_wo_it} show the radial component of the mistilt detection error (a)~without and (b)~with iterating the tilt measurement using the result of one cycle as a start for the next. Without iteration a systematic error of approximately -0.2~mrad occurs, which might already be tolerable for many applications. The measurement becomes perfectly accurate if the iterative scheme is applied, since the distribution is symmetric around zero. This behaviour of the automated evaluation procedure is independent of noise added due to a finite electron dose. Also the width of the distributions plotted in Fig.~\ref{Accuracy_w_wo_it} shows no strong dependence on electron counting noise but is reduced by the iterative scheme. A precision of about $\pm 0.13$ mrad is deduced from the range containing 68\% of the test cases, corresponding to the 1-$\sigma$ level of a normal distribution. A slight decrease  of the precision is found for the lowest probe current of 10 pA assumed in the simulations. Reduction of the incident probe current by another order of magnitude leads to a failure of the evaluation in over 50\% of the cases under the applied conditions of the test, since the low signal of the Kikuchi bands could not be distinguished from noise anymore. It should be noted though that the probe current given in this test is not a fundamental limit because the signal-to-noise ratio of the Kikuchi bands also depends on the detection dwell-time, the size and efficiency of detector pixels, as well as the number of probe positions included in a PACBED. This means that the automated tilt measurement can also be applied in cases with lower probe currents or in low-dose experiments by averaging over a larger area of the scan field and by applying pattern binning.

\section{Experimental results}
\subsection{Data acquisition}
Experimental data has been acquired with an FEI Titan 80-300~\cite{TitanS} and an FEI Titan G2 80-200~\cite{TitanChemiSTEM} electron microscope at accelerating voltages of 300 and 200~kV, respectively. The probe convergence semi-angle was limited to 25~mrad by an aperture and probe aberrations were minimized within this angular range using a double-hexapole corrector providing below 0.1 nm spatial resolution.

Momentum-resolved STEM data of a La$_{\text{0.7}}$Sr$_{\text{0.3}}$MnO$_{\text{3}}$ epitaxial layer grown on an SrTiO$_3$ substrate was recorded with a Merlin for EM Medipix~3 single chip detector (Quantum Detectors, Oxfordshire, UK) with a $256 \times 256$ pixel array \cite{Plackett2013,paterson_medipix3}. The TEM sample was prepared by focussed ion beam (FIB) milling in cross-section geometry.
A second experimental case study of high-Mn steel has been performed using an electron microscope pixel array detector (EMPAD) (Thermo Fisher Scientific) with $128 \times 128$ pixels \cite{Tate2016}. The TEM sample was prepared by a lift-out method in a dual-beam focused ion beam (FIB) FEI Helios NanoLab 460F1 \cite{HeliosFIB}.

In order to demonstrate the applicability of the method in practice, we present two case studies in the following. Firstly, it is shown that specimen bending occurring in thin foils of a perovskite heterostructure can be quantified at large fields of view (FOV). This example also shows that atomically resolved 4D-STEM with $10^6$ scan pixels is feasible despite a slow frame rate in the kHz range, i.e. for a total scan time on the order of 1000 seconds. Secondly, we consider a perfect twin boundary in a high-Mn steel specimen. Since this interface is purely crystallographic, it exhibits no compositional or strain gradients but an in-plane rotation between the adjacent grains. Observations made in this example provide information about the spatial resolution of the method.

\subsection{Bending of specimens at large fields of view}\label{sec_lmosto}

\begin{figure}
\includegraphics[width=0.98\linewidth]{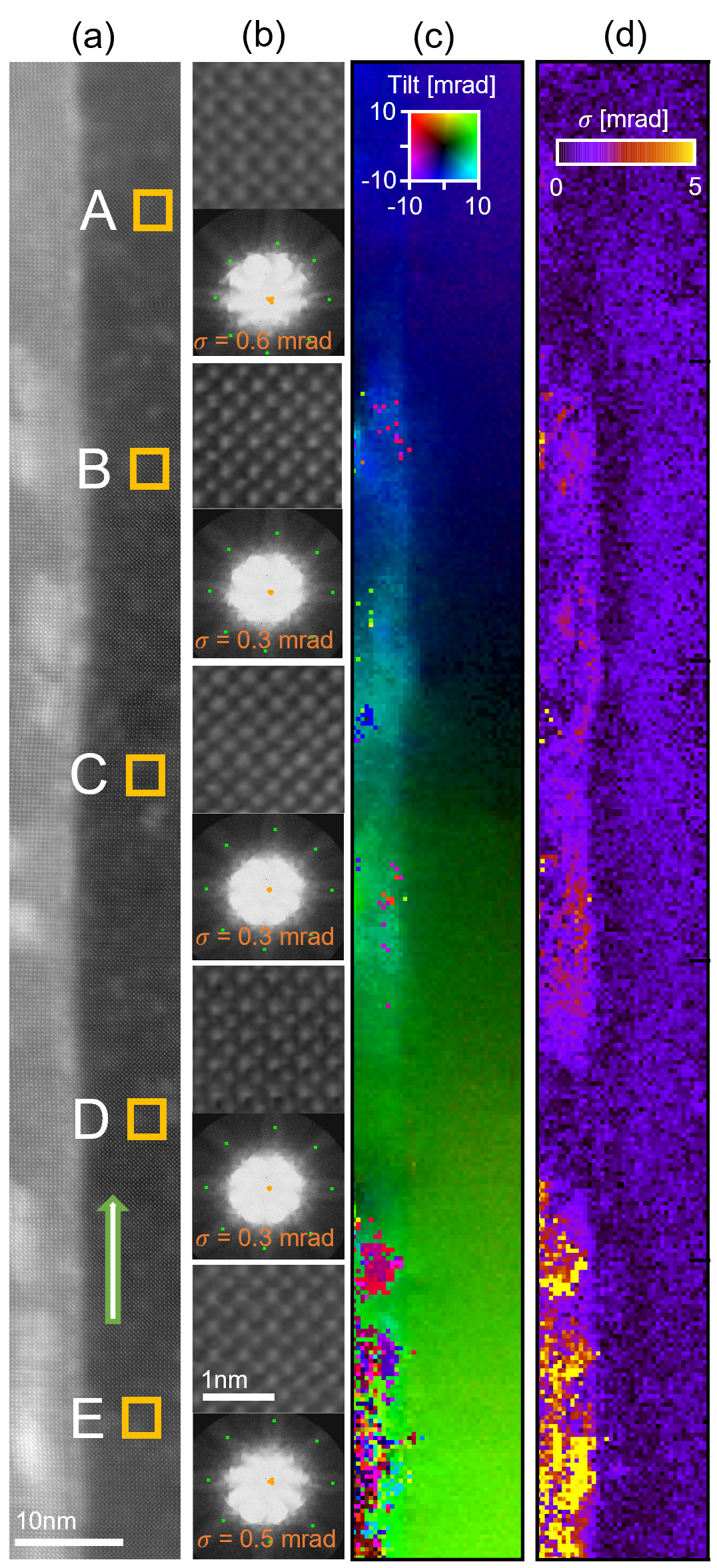}
\caption{Momentum-resolved STEM acquisition of epitaxial La$_{\text{0.7}}$Sr$_{\text{0.3}}$MnO$_{\text{3}}$ (bright layer, left) on SrTiO$_\text{3}$ (right). (a)~Virtual detector ADF image of a $17\times146\,$nm scan. (b)~Five selected areas from (a) with real space scan images and corresponding PACBED patterns. Green and yellow dots in the PACBEDs mark detected Kikuchi bands and deduced band crossings, respectively. The r.m.s. scatter $\sigma$ of crossing points is given. (c)~Map of the measured out-of-plane mistilt. (d)~Map of estimated error of the tilt measurement.}
\label{fig:large}
\end{figure}

Figure~\ref{fig:large}(a) shows the ADF signal obtained from 4D-STEM data at a scan raster of $400\times3500$, which corresponds to a field of view of 17\,nm $\times$ 146\,nm. The left-hand side contains an epitaxial La$_{\text{0.7}}$Sr$_{\text{0.3}}$MnO$_{\text{3}}$ layer (bright), grown on an SrTiO$_3$ substrate on the right hand side \cite{LMO_SMO_antwerp}. Subsets of the data are shown in Fig.~\ref{fig:large}(b) for five substrate regions A-E labelled in~(a), being approximately 30\,nm apart from each other.

Although atomic resolution is present in all ADF images, their contrast varies significantly and is maximum in regions C and D, which correspond to an approximate zone axis condition. A selection of PACBED patterns from different regions of the scan is displayed in Fig.~\ref{fig:large}\,(b). In particular, a variation of crystal tilt is identified therein by the apparent vertical shift of the crossing of the Kikuchi bands when comparing the patterns in sequence from region A to region E. In the example patterns shown, green dots mark the Kikuchi bands detected by the automated mistilt measurement. The measurement routine provides an internal error estimate $\sigma$ which corresponds to the root-mean-squared (r.m.s.) deviation of line crossings found in the analysis, e.g., as shown in Fig.~\ref{first_step_demo}. Error estimates in this experimental study are well below one milliradian, whereas the mistilt magnitude from $[100]$ zone axis reaches up to $9$\,mrad in some regions.

\begin{figure}
\includegraphics[width=1.0\linewidth]{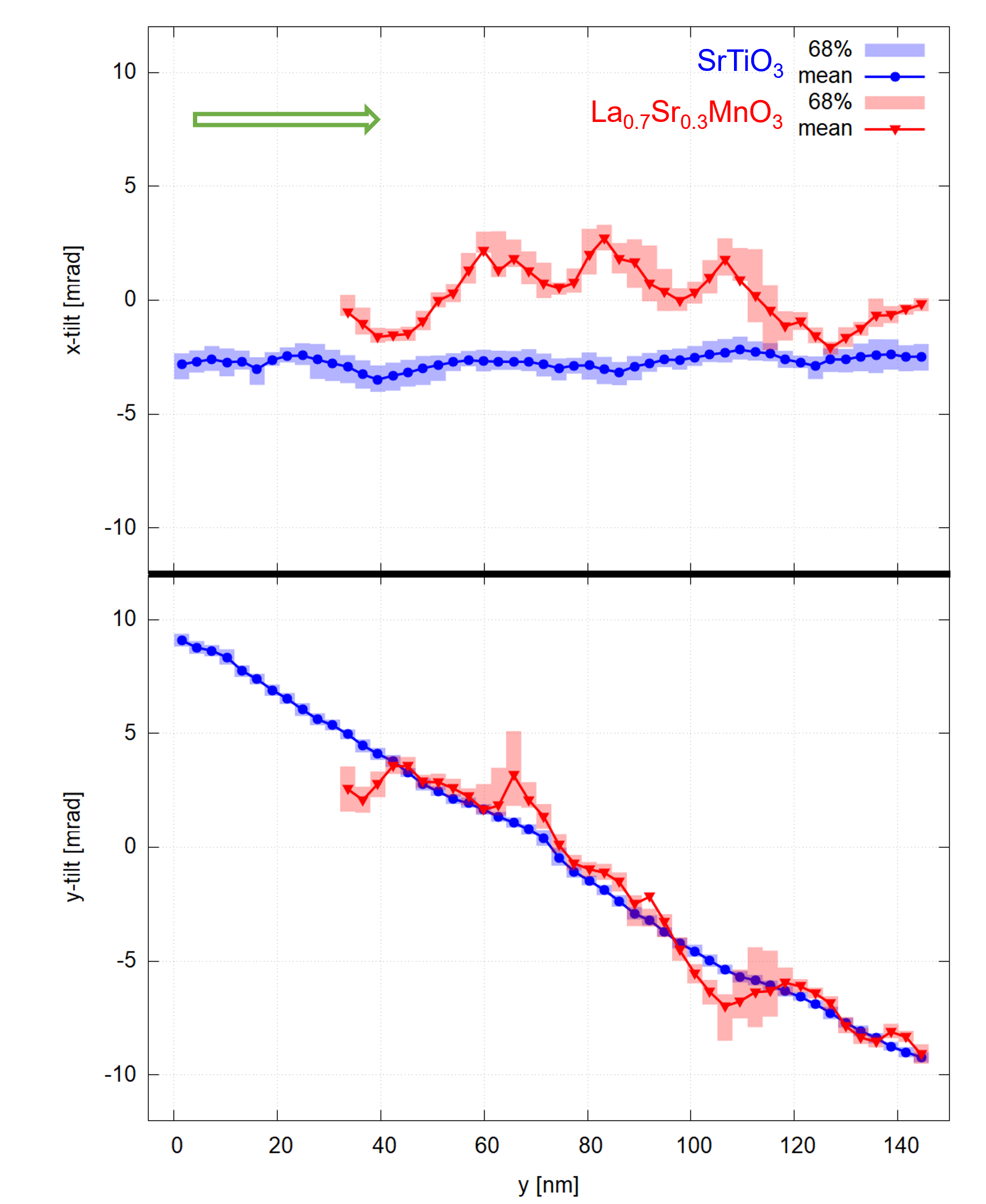}
\caption{Line profiles of sample mistilt along the arrow in Fig.~\ref{fig:large} in the epitaxial layer (red) and the substrate (blue). (a) Mistilt $x$-component, perpendicular to the interface, and (b) $y$-component, parallel to the interface, show a different variation with sample position. Whereas there is almost no change of the $x$-component, the $y$ component exhibits a linear trend, indicating sample bending of 20\,mrad over 140\,nm. Oscillations of the sample mistilt around the linear trend are observed in the epitaxial layer for both components.}
\label{sto_lmo_line}
\end{figure}

From the scan a set of $40\times350$ PACBED patterns was generated, where each PACBED is calculated from an area corresponding to one projected pseudo-cubic unit of the perovskite. The application of the iterative evaluation procedure took 20~minutes in total on a consumer-grade computer. The result of the systematic mistilt analysis for the whole region of Fig.~\ref{fig:large}\,(a) is shown in Fig.~\ref{fig:large}\,(c). A colour scale is used to display the modulus and direction of mistilt in terms of brightness and hue, respectively. The change from blue to green along the vertical scanning direction corresponds to a change of the vertical tilt component. This dominant variation of tilt is approximately a linear function of the scan position and corresponds to a bending of the specimen around a horizontal axis by approximately $\pm9\,$mrad, as plotted in Fig.~\ref{sto_lmo_line}.

Although the global trend of the tilt variation is linear, significant local deviations from this trend are observed in the epitaxial layer. These areas also appear with higher brightness in the ADF image of Fig.~\ref{fig:large}\,(a), which is indicative for enhanced static disorder due to defects. Additional diffuse intensity in the diffraction pattern due to Huang scattering \cite{Huang1947} associated with disorder hampers Kikuchi band detection by lowering their contrast. A map of the measurement error estimated by the automated procedure is plotted in Fig.~\ref{fig:large}\,(d). A good precision of 0.3 to 0.6\,mrad is obtained in the substrate while the majority of measurements in the epitaxial layer has an error below 2\,mrad. Significantly larger errors are found in some areas of the epitaxial layer, especially at the bottom of the map, where also the measured tilts scatter strongly. This indicates that the algorithm could not converge in these cases within a predefined limit of 30 iterations, most likely due to a low contrast of the Kikuchi-band pattern.

It is instructive to interpret the obtained tilt map in terms of strain-induced tilt variations caused by the epitaxial layer grown with a lattice mismatch of 0.8\,\%. For this purpose, we calculated a line profile of the mistilt along the direction indicated by the green arrow in Fig. \ref{fig:large}(a) separately for the substrate and epitaxial layer. These line profiles are plotted in Fig.~\ref{sto_lmo_line} for (a)~the component of tilt perpendicular to the interface and (b) the component of tilt parallel to the interface. The blue profile was measured from substrate areas and shows a constant bending of the lamella along the interface with a tilt gradient of 143\,$\mu$rad/nm and a standard deviation of 1\,mrad, as indicated by error bars. In contrast to that, the red curves measured on the layer material show a wave-like deviation from the linear trend of up to 7\,mrad, which are likely caused by thin foil relaxation and defects. Larger error bars of the layer measurement reflect both, the slightly worse precision of tilt detection and an increased tilt variation in the layer material also along the direction perpendicular to the interface.

\subsection{Tilt measurement in high-Mn steel}
\begin{figure}
\includegraphics[width= 0.99\linewidth]{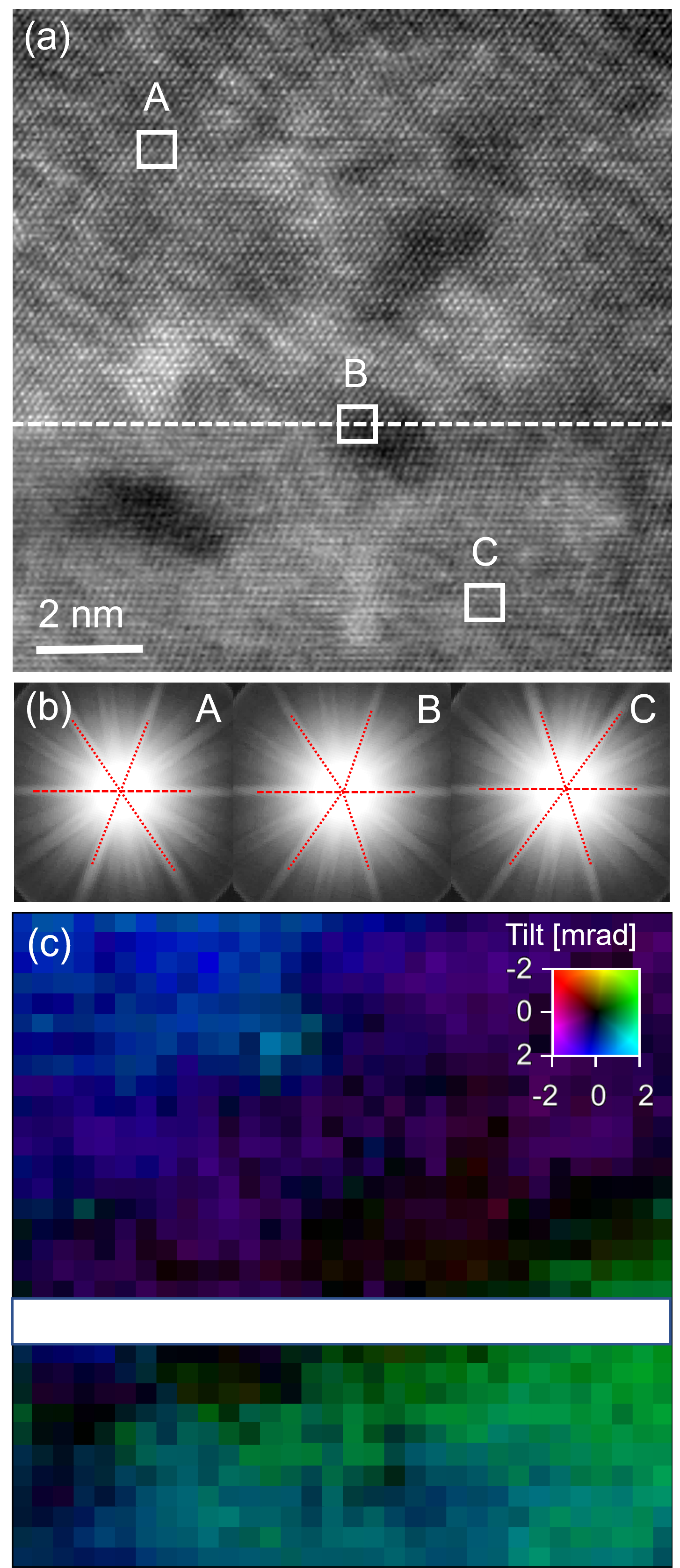} 
\caption{Mistilt mapping across a perfect twin boundary in high-Mg steel. (a)~Virtual ADF image calculated from the 4D-STEM data set. Two grains with different in-plane orientation are separated by a horizontal grain boundary. (b)~Example PACBED patterns of three selected areas from~(a) with positions A and C taken on either side of the boundary and B taken from an area that includes the boundary. (c)~Out-of-plane orientation map showing a small sample bending of approximately 2\,mrad over the entire field of view.}
\label{steel}
\end{figure}

The second case study is a tilt mapping across a twin boundary in austenitic steel. The results in Fig.~\ref{steel} show the mapping of a perfect $\Sigma$3 twin boundary viewed along the common [110] direction of the two grains with the twin boundary in a \{111\} plane and oriented horizontally in Fig.~\ref{steel}(a). From the methodological point of view, this structure is interesting since it provides an interface of a perfect twin which is well understood, however, without adding complexity due to compositional gradients or strain. Besides mapping the tilt induced by bending of the thinned TEM sample, as in the previous case, this second case provides additional insight regarding the attainable spatial resolution of the tilt measurement, because the crystallographic orientation changes abruptly at the twin boundary. The change of orientation is apparent in the PACBED patterns as an in-plane mirror with respect to the [111] interface plane, which is marked by the dashed line in Fig.~\ref{steel}(b). 

Figure~\ref{steel}(c) shows the orientation map in a scan of 14\,nm $\times$ 14\,nm. The sample appears mostly flat with tilt variations of 1 to 2\,mrad. Although these orientation changes are small, they are smooth across the twin boundary and measured with very good precision of $\sigma < 0.1$\,mrad. Above this level of uncertainty, no abrupt change of the out-of-plane crystallographic orientation was found at the grain boundary. This indicates that the two grains are indeed perfect crystallization twins.

Pattern~B displays an obvious problem for the procedure of tilt measurement at grain boundaries with significant orientation changes. There are partial bands on either side of the \{111\} plane, i.e. no counterpart is found on the opposite side to support a pair-wise relation of peaks in the azimuthal scan to a Kikuchi band. This causes the measurement routine to fail, which is reflected by the white area at the twin boundary in Fig.~\ref{steel}(c). The width of the white area corresponds to that of one unit cell on either side of the twin boundary. Already at a distance of two unit cells from the boundary, the routine works well with good accuracy and precision, i.e. the mistilt at these positions doesn't deviate much from the results obtained deeper in the grains, and the error estimates are likewise small. Accordingly, we deduce that the result in one grain is not influenced by the difference in orientation of the other grain at a distance of two unit cells, which could be interpreted as a spatial resolution of better than 0.8\,nm in this case.

\section{Discussion}
A geometric approach to measure crystal orientation from diffraction patterns has been presented. By choosing Kikuchi bands as the source of the signal for orientation mistilt measurement, the evaluation procedure requires that Kikuchi bands are separated from other much stronger features in electron diffraction patterns, such as the bright field and Bragg reflections. A sufficiently large sample thickness is required, in most cases more than 10\,nm, to generate significant Kikuchi bands at room temperature. The purpose of the method is to measure mistilt in orientations close to a high symmetry zone axis as frequently used in high-resolution STEM. In principle it is beneficial for the geometric analysis that the diffraction pattern covers a large angular range which can be achieved by choosing a short camera length. In our experiments the method produced reliable results in cases where the bright-field diameter is smaller than half the edge length of the detector. In addition, shorter camera lengths allow for larger mistilts to be analyzed and better precision. A compromise should be found when extending the angular range of detection because the signal drops towards larger angles (decay of scattering factors) and non-linear projection distortions can complicate angular calibration. Even on simulated patterns, as well as under optimized experimental conditions we found a limitation of precision of about 0.1\,mrad root-mean-square for the mistilt which is mainly due to the finite width of the Kikuchi bands.

It is instructive to study whether Kikuchi band detection is facilitated by heating specimens. Simulations show the disappearance of higher-order Laue-zone (HOLZ) rings with increasing temperature, which may otherwise lead to systematic errors or failures of the procedure. In particular, this would be the case when at least one HOLZ ring is partially but not completely included in the azimuthal scan. The diffuse scattering signal increases with increasing sample temperature on the one hand. However, the simulations do not show a likewise increased contrast of Kikuchi bands on the other hand. Apart from the disappearance of the HOLZ signal, heating the sample does not seem to offer significant advantages. 

At lower beam energies and large mistilt Kikuchi bands can appear curved in the diffraction pattern, since their edges follow conic sections in the strict sense. As a consequence, the prediction of the crossing point of the bands with the present method could become slightly inaccurate. In this case, a final step evaluating the full two-dimensional course of the Kikuchi bands could be necessary, e.g., assuming hyperbolic shapes. However, in the cases presented here, sufficiently high beam energies have been used such that this error is negligible.

\begin{figure}
\includegraphics[width=0.95\linewidth]{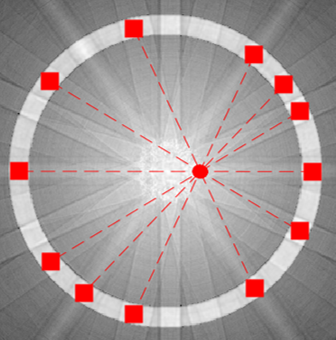}
\caption{Simulated PACBED pattern of Fe [110] fcc steel and example of a false detection of the Kikuchi band intersection with a very small root-mean-square deviation. This is a rare case, where by chance, the depicted wrong association of pairs of peaks to bands leads to the smallest deviation of common intersections. Given the underlying 2-dimensional pattern, the error is obvious, but not detectable in the evaluation of a single azimuthal scan. Using the iterative scheme, this false detection is nevertheless eliminated.}
\label{bad_luck}
\end{figure}

A significant step in the pattern analysis is reducing a two-dimensional problem, the intersection of bands, to a one-dimensional problem, measuring peak positions in an azimuthal intensity profile. This is beneficial for the speed and complexity of the numerical evaluation. However, a problem may arise when associating pairs of peaks to bands. The smallest root-mean-square distance of band intersections is used as the main criterion for finding the correct association. There are rare cases, even under good experimental conditions, where this criterion yields a deceptively good but actually wrong result. An example is depicted in Fig.~\ref{bad_luck}. Note that the computer only operates based on the 1D scan and doesn’t perceive the 2D pattern like a human observer. These initially wrong results are, however, healed by the refinement step using another azimuthal scan extracted from a different area of the 2D pattern. In all cases observed so far, this procedure converged to the correct result.

Applying the method to 4D-STEM data allows for mapping local mistilt. The two experimental case studies presented in the results section both show systematic variations in local orientation. Each pixel in the maps corresponds to a scanned area from which diffraction patterns have been averaged and then analyzed independently. The locality of tilt information, however, needs to be interpreted with care due to the significant thickness- and orien\-ta\-tion-dependent broadening of the electron probe in the specimen. An empirical assessment of the spatial resolution for mistilt variation is found in the mapping of Fig.~\ref{steel} for a perfect twin-boundary in steel. Already at a distance of one unit cell from the twin boundary to either side, the measured tilts are consistent with the rest of the grain, which suggests that the mistilt information is generated from a projected area with a radius of approximately one unit cell in the present case. For patterns generated from areas closer to the grain boundary, e.g. pattern B in Fig.~\ref{steel}(b), the method fails for obvious reasons. In this case all Kikuchi bands except one terminate at the pole and do not generate the expected signal on the opposite side of the azimuthal scan. In the case of high-resolution STEM of steel, a continuous Kikuchi band along a direction is therefore generated as soon as there is at least one atomic column from the scan position in both directions to form a crystal plane. For the most intense Kikuchi bands, the neighboring columns have distances of one or two unit cells. The spatial resolution for local tilt mapping is therefore determined by the spacing of atomic columns in the crystallographic planes corresponding to the Kikuchi bands used in the analysis. Since all orientation mappings presented in this study were generated from PACBED patterns of at least unit cell areas, tilt information in one data point can be seen as independent from its neighbors.

\section{Summary and Conclusion}

To measure small tilts from the zone axis geometry, an approach based on the signal from Kikuchi bands in diffraction patterns was chosen. This approach requires a minimum of a-priori information and can thus be used with little preliminary work. The need to record Kikuchi bands does not preclude simultaneous evaluation of bright field or elastic diffraction signal as long as both are recorded with sufficient area on the detector. This means that, for example, with a momentum-resolved STEM measurement (4D-STEM), data relevant to material science as well as supporting data important for later modelling, such as tilt, can be recorded simultaneously.

The evaluation procedure has been implemented numerically and provides fast and reliable results with accuracy on the order of 0.1 mrad mistilt. However, the application of the method to typical (S)TEM diffraction patterns is limited to small mistilts of less than 5° and requires a crystal thickness sufficient to form Kikuchi bands. As a further development it could be investigated whether the need for minimal inputs can be completely avoided, e.g. by training a neural network with multiple patterns, to obtain a fully automatic method.

Two examples have demonstrated that the method can be used to map sample mistilts over large areas and interfaces. The resulting information allows for the measurement of orientation changes across interfaces and to determine grain boundary parameters. This is important in modelling, e.g., when diffraction effects need to be separated from that of chemical gradients. The high speed of the evaluation allows fast feedback for an online sample-tilt optimization. This would be a step towards automating high resolution transmission electron microscopy in finding zone-axis orientation or applying deliberate mistilts for, e.g., quantitative EDX.

\section{Acknowledgements}
M.C. acknowledges funding from the J\"{u}lich Melbourne Postgraduate Academy (JUMPA). K.M.-C. acknowledges funding from the Deutsche Forschungsgemeinschaft (DFG) under grant number EXC 2089/1 - 390776260 (Germany's Excellence Strategy) and the Helmholtz association under contract VH-NG-1317. J.V. and K.M.-C. received funding from the FWO under contract G042920N. M.L.-C acknowledges financial support of the DFG  within the Collaborative Research Center (SFB) 761 “Steel – ab initio". The collection of EMPAD data took place within a cooperation with Thermo Fisher Scientific which is gratefully acknowledged.

\appendix
\section{Calculation of line crossings from positions in an azimuthal scan}
\label{AppCalcCrossing}
Given is a list of $2n$ unequal azimuthal angles $\varphi_i$ corresponding to positions
\begin{equation}
    {\bf X}_i = {\bf Q} + r \begin{pmatrix} \cos \varphi_i \\ \sin \varphi_i \end{pmatrix}
    \label{eq:a1}
\end{equation}
on a circle of radius $r$ with center $\bf Q$ in a two-dimensional plane. The list is sorted, such that $\varphi_i < \varphi_{i+1}$ for all $i \geq 1$ and $i < 2n.$ $n$ lines are constructed by connecting pairs of positions ${\bf X}_i$ and ${\bf X}_{i+n}$ for $i \leq n.$ A line that contains two points ${\bf X}_i$ and ${\bf X}_j$ is given by the formula
\begin{equation}
    {\bf L}_{ij}(t) = {\bf X}_i + t ({\bf X}_{j} - {\bf X}_i), t \in \mathbb{R}.
    \label{eq:a2}
\end{equation}
The crossing point of two different lines ${\bf L}_{ij}(t_1)$ and ${\bf L}_{kl}(t_2)$ with $k \neq i,$ $j = i + n,$ and $l = k + n$ is found by solving the system of two equations given by
\begin{equation}
    {\bf 0} = {\bf L}_{ij}(t_1) - {\bf L}_{kl}(t_2)
    \label{eq:a3}
\end{equation}
for the parameters $t_1$ and $t_2$ and calculating the respective position via Eq.~(\ref{eq:a2}). The solution is found as
\begin{equation}
    \begin{pmatrix} t_1 \\ t_2 \end{pmatrix} = A^{-1} {\bf B},
    \label{eq:a4}
\end{equation}
with the $2 \times 2$ matrix $A$ given by
\begin{equation}
    A = \left( {\bf X}_j - {\bf X}_i \ {\bf X}_k - {\bf X}_l \right)
    \label{eq:a5}
\end{equation}
and the right side vector
\begin{equation}
    {\bf B} = {\bf X}_k - {\bf X}_i.
    \label{eq:a6}
\end{equation}


\begin{thebibliography}{10}
\expandafter\ifx\csname url\endcsname\relax
  \def\url#1{\texttt{#1}}\fi
\expandafter\ifx\csname urlprefix\endcsname\relax\def\urlprefix{URL }\fi
\expandafter\ifx\csname href\endcsname\relax
  \def\href#1#2{#2} \def\path#1{#1}\fi

\bibitem{pennycook1988zcontrast}
S.~J. Pennycook, L.~A. Boatner, Chemically sensitive structure-imaging with a
  scanning transmission electron microscope, Nature 336 (1988) 565--567.
\newblock \href {https://doi.org/10.1038/336565a0}
  {\path{doi:10.1038/336565a0}}.

\bibitem{findlay2010dynamics}
S.~D. Findlay, N.~Shibata, H.~Sawada, E.~Okunishi, Y.~Kondo, Y.~Ikuhara,
  Dynamics of annular bright field imaging in scanning transmission electron
  microscopy, Ultramicroscopy 110 (2010) 903--923.
\newblock \href {https://doi.org/10.1016/j.ultramic.2010.04.004}
  {\path{doi:10.1016/j.ultramic.2010.04.004}}.

\bibitem{bosman2007atomiceels}
M.~Bosman, V.~J. Keast, J.~L. Garcia-Munoz, A.~J. D’Alfonso, S.~D. Findlay,
  L.~J. Allen, Two-dimensional mapping of chemical information at atomic
  resolution, Physical Review Letters 99~(8) (2007) 086102.
\newblock \href {https://doi.org/10.1103/PhysRevLett.99.086102}
  {\path{doi:10.1103/PhysRevLett.99.086102}}.

\bibitem{dalfonso2010atomicedx}
A.~J. {D’Alfonso}, B.~Freitag, D.~Klenov, L.~J. Allen, Atomic-resolution
  chemical mapping using energy-dispersive {X}-ray spectroscopy, Physical
  Review B 81~(10) (2010) 100101.
\newblock \href {https://doi.org/10.1103/PhysRevB.81.100101}
  {\path{doi:10.1103/PhysRevB.81.100101}}.

\bibitem{maccagnano2008effects}
S.~E. Maccagnano-Zacher, K.~A. Mkhoyan, E.~J. Kirkland, J.~Silcox, Effects of
  tilt on high-resolution {ADF-STEM} imaging, Ultramicroscopy 108~(8) (2008)
  718--726.
\newblock \href {https://doi.org/10.1016/j.ultramic.2007.11.003}
  {\path{doi:10.1016/j.ultramic.2007.11.003}}.

\bibitem{Zhou2016}
D.~Zhou, K.~M{\"u}ller-Caspary, W.~Sigle, F.~F. Krause, A.~Rosenauer, P.~A.
  {van Aken}, Sample tilt effects on atom column position determination in
  {ABF–STEM} imaging, Ultramicroscopy 160 (2016) 110--117.
\newblock \href {https://doi.org/10.1016/j.ultramic.2015.10.008}
  {\path{doi:10.1016/j.ultramic.2015.10.008}}.

\bibitem{macarthur2021optimizing}
K.~E. {MacArthur}, A.~B. Yankovich, A.~B{\'e}ch{\'e}, M.~Luysberg, H.~G. Brown,
  S.~D. Findlay, M.~Heggen, L.~J. Allen, Optimizing experimental conditions for
  accurate quantitative energy-dispersive {X}-ray analysis of interfaces at the
  atomic scale, Microscopy and Microanalysis 27 (2021) 528--542.
\newblock \href {https://doi.org/10.1017/S1431927621000246}
  {\path{doi:10.1017/S1431927621000246}}.

\bibitem{MacArthur2017}
K.~E. MacArthur, H.~G. Brown, S.~D. Findlay, L.~J. Allen, Probing the effect of
  electron channelling on atomic resolution energy dispersive {X}-ray
  quantification, Ultramicroscopy 182 (2017) 264--275.
\newblock \href {https://doi.org/10.1016/j.ultramic.2017.07.020}
  {\path{doi:10.1016/j.ultramic.2017.07.020}}.

\bibitem{Lugg2015}
N.~R. Lugg, G.~Kothleitner, N.~Shibata, Y.~Ikuhara, On the quantitativeness of
  {EDS STEM}, Ultramicroscopy 151 (2015) 150--159.
\newblock \href {https://doi.org/10.1016/j.ultramic.2014.11.029}
  {\path{doi:10.1016/j.ultramic.2014.11.029}}.

\bibitem{Chen2021}
Z.~Chen, Y.~Jiang, Y.-T. Shao, M.~E. Holtz, M.~Odstr{\v c}il,
  M.~Guizar-Sicairos, I.~Hanke, S.~Ganschow, D.~G. Schlom, D.~A. Muller,
  \href{https://science.sciencemag.org/content/372/6544/826}{Electron
  ptychography achieves atomic-resolution limits set by lattice vibrations},
  Science 372~(6544) (2021) 826--831.
\newblock \href
  {http://arxiv.org/abs/https://science.sciencemag.org/content/372/6544/826.full.pdf}
  {\path{arXiv:https://science.sciencemag.org/content/372/6544/826.full.pdf}},
  \href {https://doi.org/10.1126/science.abg2533}
  {\path{doi:10.1126/science.abg2533}}.

\bibitem{Sha2023}
H.~Sha, Y.~Ma, G.~Cao, J.~Cui, W.~Yang, Q.~Li, R.~Yu, {Sub-nanometer-scale
  mapping of crystal orientation and depth-dependent structure of dislocation
  cores in SrTiO$_3$}, Nature Communications 14~(1) (2023) 162.
\newblock \href {https://doi.org/10.1038/s41467-023-35877-7}
  {\path{doi:10.1038/s41467-023-35877-7}}.

\bibitem{Diederichs2024}
B.~Diederichs, Z.~Herdegen, A.~Strauch, F.~Filbir, K.~M{\"u}ller-Caspary,
  \href{https://doi.org/10.1038/s41467-023-44268-x}{Exact inversion of
  partially coherent dynamical electron scattering for picometric structure
  retrieval}, Nature Communications 15~(1) (2024) 101.
\newblock \href {https://doi.org/10.1038/s41467-023-44268-x}
  {\path{doi:10.1038/s41467-023-44268-x}}.

\bibitem{Broek2012}
W.~Van~den Broek, C.~T. Koch,
  \href{https://link.aps.org/doi/10.1103/PhysRevLett.109.245502}{Method for
  retrieval of the three-dimensional object potential by inversion of dynamical
  electron scattering}, Physical Review Letters 109 (2012) 245502.
\newblock \href {https://doi.org/10.1103/PhysRevLett.109.245502}
  {\path{doi:10.1103/PhysRevLett.109.245502}}.

\bibitem{liao2018}
Z.~Liao, N.~Gauquelin, R.~J. Green, K.~Müller-Caspary, I.~Lobato, L.~Li, S.~V.
  Aert, J.~Verbeeck, M.~Huijben, M.~N. Grisolia, V.~Rouco, R.~{El Hage}, J.~E.
  Villegas, A.~Mercy, M.~Bibes, P.~Ghosez, G.~A. Sawatzky, G.~Rijnders,
  G.~Koster, Metal-insulator-transition engineering by modulation tilt-control
  in perovskite nickelates for room temperature optical switching, Proceedings
  of the National Academy of Sciences 115~(38) (2018) 9515--9520.
\newblock \href {https://doi.org/10.1073/pnas.1807457115}
  {\path{doi:10.1073/pnas.1807457115}}.

\bibitem{calcagnotto2010}
M.~Calcagnotto, D.~Ponge, E.~Demir, D.~Raabe, Orientation gradients and
  geometrically necessary dislocations in ultrafine grained dual-phase steels
  studied by 2d and 3d {EBSD}, Materials Science and Engineering: A 527~(10)
  (2010) 2738--2746.
\newblock \href {https://doi.org/https://doi.org/10.1016/j.msea.2010.01.004}
  {\path{doi:https://doi.org/10.1016/j.msea.2010.01.004}}.

\bibitem{Strauch2023}
A.~Strauch, B.~M{\"a}rz, T.~Denneulin, M.~Cattaneo, A.~Rosenauer,
  K.~M{\"u}ller-Caspary,
  \href{https://doi.org/10.1093/micmic/ozad016}{{Systematic errors of electric
  field measurements in ferroelectrics by unit cell averaged momentum transfers
  in STEM}}, Microscopy and Microanalysis 29 (2023) 499--511.
\newblock \href {https://doi.org/10.1093/micmic/ozad016}
  {\path{doi:10.1093/micmic/ozad016}}.

\bibitem{Muller2012a}
K.~M{\"u}ller, H.~Ryll, I.~Ordavo, S.~Ihle, L.~Str{\"u}der, K.~Volz, J.~Zweck,
  H.~Soltau, A.~Rosenauer, Scanning transmission electron microscopy strain
  measurement from millisecond frames of a direct electron charge coupled
  device, Applied Physics Letters 101 (2012) 212110.
\newblock \href {https://doi.org/10.1063/1.4767655}
  {\path{doi:10.1063/1.4767655}}.

\bibitem{Plackett2013}
R.~Plackett, I.~Horswell, E.~N. Gimenez, J.~Marchal, D.~Omar, N.~Tartoni,
  Merlin: a fast versatile readout system for {Medipix3}, Journal of
  Instrumentation 8 (2013) C01038.
\newblock \href {https://doi.org/10.1088/1748-0221/8/01/C01038}
  {\path{doi:10.1088/1748-0221/8/01/C01038}}.

\bibitem{Muller-Caspary2015a}
K.~M{\"u}ller-Caspary, A.~Oelsner, P.~Potapov, Two-dimensional strain mapping
  in semiconductors by nano-beam electron diffraction employing a delay-line
  detector, Applied Physics Letters 107 (2015) 072110.
\newblock \href {https://doi.org/10.1063/1.4927837}
  {\path{doi:10.1063/1.4927837}}.

\bibitem{Ryll2016}
H.~Ryll, M.~Simson, R.~Hartmann, P.~Holl, M.~Huth, S.~Ihle, Y.~Kondo,
  P.~Kotula, A.~Liebel, K.~M{\"u}ller-Caspary, A.~Rosenauer, R.~Sagawa,
  J.~Schmidt, H.~Soltau, L.~Str{\"u}der, A {pnCCD}-based, fast direct single
  electron imaging camera for {TEM} and {STEM}, Journal of Instrumentation 11
  (2016) P04006.
\newblock \href {https://doi.org/10.1088/1748-0221/11/04/P04006}
  {\path{doi:10.1088/1748-0221/11/04/P04006}}.

\bibitem{Tate2016}
M.~W. Tate, P.~Purohit, D.~Chamberlain, K.~X. Nguyen, R.~Hovden, C.~S. Chang,
  P.~Deb, E.~Turgut, J.~T. Heron, D.~G. Schlom, D.~C. Ralph, G.~D. Fuchs, K.~S.
  Shanks, H.~T. Philipp, D.~A. Muller, S.~M. Gruner, High dynamic range pixel
  array detector for scanning transmission electron microscopy, Microscopy
  Microanalysis 22 (2016) 237--249.
\newblock \href {https://doi.org/10.1017/S1431927615015664}
  {\path{doi:10.1017/S1431927615015664}}.

\bibitem{Jannis2022}
D.~Jannis, C.~Hofer, C.~Gao, X.~Xie, A.~Béché, T.~J. Pennycook, J.~Verbeeck,
  Event driven {4D STEM} acquisition with a {Timepix3} detector: {M}icrosecond
  dwell time and faster scans for high precision and low dose applications,
  Ultramicroscopy 233 (2022) 113423.
\newblock \href {https://doi.org/10.1016/j.ultramic.2021.113423}
  {\path{doi:10.1016/j.ultramic.2021.113423}}.

\bibitem{xu_lebeau_cnn}
W.~Xu, J.~M. LeBeau, A deep convolutional neural network to analyze position
  averaged convergent beam electron diffraction patterns, Ultramicroscopy 188
  (2018) 59--69.
\newblock \href {https://doi.org/10.1016/j.ultramic.2018.03.004}
  {\path{doi:10.1016/j.ultramic.2018.03.004}}.

\bibitem{lebeau_pacbed_2010}
J.~M. LeBeau, S.~D. Findlay, L.~J. Allen, S.~Stemmer, Position averaged
  convergent beam electron diffraction: {T}heory and applications,
  Ultramicroscopy 110 (2010) 118--125.
\newblock \href {https://doi.org/10.1016/j.ultramic.2009.10.001}
  {\path{doi:10.1016/j.ultramic.2009.10.001}}.

\bibitem{nishikawa1928diffraction}
S.~Nishikawa, S.~Kikuchi, Diffraction of cathode rays by mica, Nature
  121~(3061) (1928) 1019--1020.
\newblock \href {https://doi.org/10.1038/1211019a0}
  {\path{doi:10.1038/1211019a0}}.

\bibitem{Venables1973}
J.~A. Venables, C.~J. Harland, Electron back-scattering patterns — {A} new
  technique for obtaining crystallographic information in the scanning electron
  microscope, The Philosophical Magazine: A Journal of Theoretical Experimental
  and Applied Physics 27~(5) (1973) 1193--1200.
\newblock \href {https://doi.org/10.1080/14786437308225827}
  {\path{doi:10.1080/14786437308225827}}.

\bibitem{trimby_kik_sem}
P.~W. Trimby, Orientation mapping of nanostructured materials using
  transmission {Kikuchi} diffraction in the scanning electron microscope,
  Ultramicroscopy 120 (2012) 16--24.
\newblock \href {https://doi.org/10.1016/j.ultramic.2012.06.004}
  {\path{doi:10.1016/j.ultramic.2012.06.004}}.

\bibitem{burton_Kik_EDAX}
G.~L. Burton, S.~Wright, A.~Stokes, D.~R. Diercks, A.~Clarke, B.~P. Gorman,
  Orientation mapping with kikuchi patterns generated from a focused stem probe
  and indexing with commercially available {EDAX} software, Ultramicroscopy 209
  (2020) 112882.
\newblock \href {https://doi.org/10.1016/j.ultramic.2019.112882}
  {\path{doi:10.1016/j.ultramic.2019.112882}}.

\bibitem{reimer_tem}
L.~Reimer, H.~Kohl, {T}ransmission {E}lectron {M}icroscopy, Springer, New York,
  NY, 2008.
\newblock \href {https://doi.org/10.1007/978-0-387-40093-8}
  {\path{doi:10.1007/978-0-387-40093-8}}.

\bibitem{cowley1957scattering}
J.~M. Cowley, A.~F. Moodie, The scattering of electrons by atoms and crystals.
  {I}. {A} new theoretical approach, Acta Crystallographica 10~(10) (1957)
  609--619.
\newblock \href {https://doi.org/10.1107/S0365110X57002194}
  {\path{doi:10.1107/S0365110X57002194}}.

\bibitem{barthel_drprobe_2018}
J.~Barthel, {D}r. {P}robe: {A} software for high-resolution {STEM} image
  simulation, Ultramicroscopy 193 (2018) 1--11.
\newblock \href {https://doi.org/10.1016/j.ultramic.2018.06.003}
  {\path{doi:10.1016/j.ultramic.2018.06.003}}.

\bibitem{Forbes2010}
B.~D. Forbes, A.~V. Martin, S.~D. Findlay, A.~J. {D'Alfonso}, L.~J. Allen,
  Quantum mechanical model for phonon excitation in electron diffraction and
  imaging using a {B}orn-{O}ppenheimer approximation, Physical Review B 82
  (2010) 104103.
\newblock \href {https://doi.org/10.1103/PhysRevB.82.104103}
  {\path{doi:10.1103/PhysRevB.82.104103}}.

\bibitem{straumanis1969lattice}
M.~E. Straumanis, D.~C. Kim, Lattice constants, thermal expansion coefficients,
  densities and perfection of structure of pure iron and of iron loaded with
  hydrogen, International Journal of Materials Research 60~(4) (1969) 272--277.
\newblock \href {https://doi.org/10.1515/ijmr-1969-600404}
  {\path{doi:10.1515/ijmr-1969-600404}}.

\bibitem{Peng_DW_factors}
L.-M.Peng, G.~Ren, S.~L. Dudarev, M.~J. Whelan, {Debye{--}Waller Factors and
  Absorptive Scattering Factors of Elemental Crystals}, Acta Crystallographica
  Section A 52~(3) (1996) 456--470.
\newblock \href {https://doi.org/10.1107/S010876739600089X}
  {\path{doi:10.1107/S010876739600089X}}.

\bibitem{TitanS}
\mbox{Ernst Ruska-Centre for Microscopy and Spectroscopy with} \mbox{Electrons
  (ER-C)}, Fei titan 80-300 stem, Journal of large-scale research facilities 2
  (2016) A42.
\newblock \href {https://doi.org/10.17815/jlsrf-2-67}
  {\path{doi:10.17815/jlsrf-2-67}}.

\bibitem{TitanChemiSTEM}
\mbox{Ernst Ruska-Centre for Microscopy and Spectroscopy with} \mbox{Electrons
  (ER-C)}, Fei titan g2 80-200 crewley, Journal of large-scale research
  facilities 2 (2016) A43.
\newblock \href {https://doi.org/10.17815/jlsrf-2-68}
  {\path{doi:10.17815/jlsrf-2-68}}.

\bibitem{paterson_medipix3}
G.~W. Paterson, R.~J. Lamb, R.~Ballabriga, D.~Maneuski, V.~{O'Shea},
  D.~{McGrouther}, Sub-100 nanosecond temporally resolved imaging with the
  {M}edipix3 direct electron detector, Ultramicroscopy 210 (2020) 112917.
\newblock \href {https://doi.org/10.1016/j.ultramic.2019.112917}
  {\path{doi:10.1016/j.ultramic.2019.112917}}.

\bibitem{HeliosFIB}
M.~Kruth, D.~Meertens, K.~Tillmann, {FEI} helios nanolab 460f1 fib-sem, Journal
  of large-scale research facilities JLSRF 2 (03 2016).
\newblock \href {https://doi.org/10.17815/jlsrf-2-105}
  {\path{doi:10.17815/jlsrf-2-105}}.

\bibitem{LMO_SMO_antwerp}
M.~Keunecke, F.~Lyzwa, D.~Schwarzbach, V.~Roddatis, N.~Gauquelin,
  K.~M{\"u}ller-Caspary, J.~Verbeeck, S.~J. Callori, F.~Klose, M.~Jungbauer,
  V.~Moshnyaga, High-$t_\mathrm{C}$ interfacial ferromagnetism in
  srmno$_3$/lamno$_3$ superlattices, Advanced Functional Materials 30~(18)
  (2020) 1808270.
\newblock \href {https://doi.org/10.1002/adfm.201808270}
  {\path{doi:10.1002/adfm.201808270}}.

\bibitem{Huang1947}
K.~Huang, {X}-ray reflexions from dilute solid solutions, Proc. Roy. Soc. Lond.
  A 190~(1020) (1947) 102--117.
\newblock \href {https://doi.org/10.1098/rspa.1947.0064}
  {\path{doi:10.1098/rspa.1947.0064}}.

\end{thebibliography}

\end{document}